# Photonic Hook with Modulated Bending Angle Formed by Using Triangular Mesoscale Janus Prisms

Wei-Yu Chen [1], Cheng-Yang Liu [1,2], Yu-Kai Hsieh [1], Oleg V. Minin [3] and Igor V. Minin [3,*]

[1] Department of Biomedical Engineering, National Yang Ming Chiao Tung University, Taipei 112, Taiwan; sam30261014.be09@nycu.edu.tw (W.-Y.C.), robin88041988@gmail.com (Y.-K.H.)

[2] Medical Device Innovation and Translation Center, National Yang Ming Chiao Tung University, Taipei City 11221, Taiwan; cyliu66@nycu.edu.tw (C.-Y.L.)

[3] Nondestructive testing school, Tomsk Polytechnic University, 30 Lenin Ave., Tomsk 634050, Russia; prof.minin@gmail.com (O.V.M.)

* Correspondence: ivminin@tpu.ru

**Abstract:** In this study, we propose a novel design of the triangular mesoscale Janus prisms for the generation of the long photonic hook. The numerical simulations based on the finite-difference time-domain method are used to examine the formation mechanism of the photonic hook. The electric intensity distributions near the micro-prisms are calculated for operating at different refractive indices and spaces of the two triangular micro-prisms. The asymetric vortexes of intensity distributions result in the long photonic hook with large bending angle. The length and the bending angle of the photonic hook are efficiently modulated by changing the space between the two triangular micro-prisms. Moreover, the narrow width of the photonic hook is achieved beyond the diffraction limit. The triangular Janus micro-prisms have high potential for practical applications in optical tweezers, nanoparticle sorting and manipulation and photonic circuits.

## 1. Introduction

Pyramidal structures, which are drawn from the Greek words *Pyro* (fire) and *Amid* (from the centre), are one of the simple geometric shapes widely found in nature. Many molecules and crystals have the shape of a pyramid [1-3]. It is known that the space within the pyramids generates or/and enhances energy in the electro-magnetic band [4-6]. The pyramidal shapes are used in three-fold rotational symmetry quantum dots [7], to enhance the light-capturing ability in sensors [8], in nano- and meso-scale resonators with high Q-factor [9,10], in Si-based photodetectors [11,12] and solar cell [13,14], for subwavelength light focusing [15,16], to form Bessel beam [17], to enhance Raman scattering [18], in food and health technologies [19,20]. Recently, it was shown that to form the photonic jets (PJ) [21,22] the pyramidal structures can be used for enhancing a focused optical field [22-26]. For example, the micropyramid array enhances the interference effect of incident and scattered lights, and the intensity of the focused field reaches 33.8-times that of the incident light [18]. Photonic hook (PH) is a new type of the PJ that the artificial curved beam focused by a Janus dielectric particle with a waist less than the half of wavelength [27,28]. The PH forming mechanism requires asymmetry of illumination wave front, the dielectric particle in the manner of geometric shape, or the optical properties of particle material [29,30]. Additionally, double PHs can be formed using two coherent illuminations [31], adjacent dielectric cylinders [32], or twin-ellipse mesoscale cylinders [33]. The PH has the potential to revolutionize mesotronics [34,35] in wide fields of applications, including optical trapping, subwavelength imaging, and signal switching. However, despite the abundance of options for obtaining PH, the PH based on pyramidal particles has not yet been considered. Obviously, the PH properties of the mesoscale Janus particles based on pyramids are worth further investigation when the multi-dielectric structures are considered [36,37]. The main purpose of the article is to identify the key characteristics of the PH based on Janus particles from triangular prisms. The possibility of generating long PHs makes it possible to expand the arsenal of methods for creating structured localized beams of this type and related applications.

## 2. Simulation model

In this study, the physical mechanism of the near-field spatial intensity distributions of optical scattering is considered by the triangular Janus micro-prisms. The optical diffraction in the Janus prisms with micro-scale dimensions is a near-field problem because of the interference by diffracted and scattered lights on an inhomogeneous medium. Since such light scattering problem has no analytical solution, the numerical approach is only suitable for this complicated inhomogeneous material. Figure 1 shows the conception of curved optical focusing beam by means of an index-contrast triangular Janus micro-prism. The width and height of the triangular micro-prism are $w$ = 6 μm and $h$. The refractive indices of the two triangular micro-prisms are $n_1$ and $n_2$. The space between the two triangular micro-prisms is $d$. The triangular Janus micro-prism is surrounded by air ($n_0$ = 1). The laser beam with 671-nm wavelength with linear polarization along $z$-axis is introduced into the bottom of the triangular Janus micro-prism along positive $x$ direction. The photonic hook along positive $x$ direction is generated by the triangular Janus micro-prism. For characterizing the photonic hook, the focal length between the point of maximum intensity peak and the vertex of the triangular prism along the $x$ direction is $f$. The bending angle of the photonic hook is $\theta$, which is relative to the incident direction of light beam and the end point at the field decayed to $1/e$ of the maximum intensity peak [28]. To demonstrate the performance of the photonic hook, a two-dimensional finite-difference time-domain computation is implemented with the perfectly matched absorbing boundaries [38]. The computational field is in $x$-$y$ plane, and the triangular prism along $z$ direction is regarded as infinite length. The triangular grid mesh is used to ensure the accuracy and speed of numerical calculation, which is set as 10 nm in the computational field.

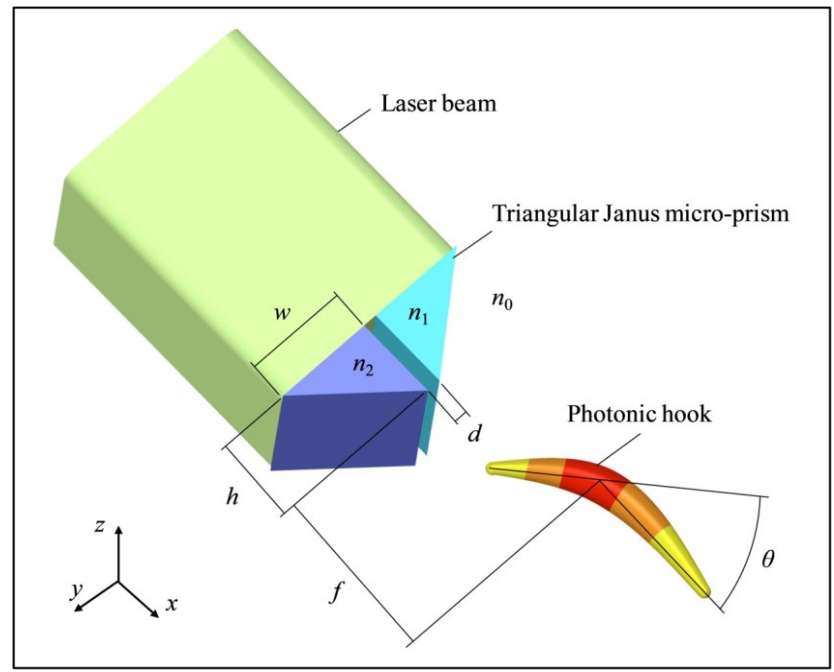

**Figure 1.** Schematic stereogram of the triangular Janus micro-prism for photonic hook.

## 3. Simulations and Results

First, the formations of classical PJ by the conventional triangular all dielectric prisms with different heights $h$ are shown in Figure 2. When the height of the mi-cro-prism decreases, the focus point of the electromagnetic field moves away from the micro-prisms and the intensity at the focus point decreases. The position of the maximum intensity field is crucial to the length of the PJ. A decrease in h as the PJ length increases is caused by the movement of the focus point outside the micro-prisms. The key parameters of the PJ are the maximum intensity enhancement, the full width at half maximum (FWHM), and the focal distance from the shadow surface of Janus particle to the point with maximal field intensity. One can see from Figure 2 that with increasing of the height of a prism the maximal field intensity on PJ increase, but the length of the PJ, FWHM and the focal distance decrease. In quantitative terms, the corresponding dependencies are shown in Figure 3. Let's note that the waist of the PJ decreases and tends to be smaller than half wavelength, while the height $h$ decreases. The height $h$ leads to an improvement in the key parameters of the PJ, which is because height $h$ increases the field intensity along the beam propagation direction. It could be noted that FWHM of the PJ is always subwavelength and at $h > 2.25$ it is less than the diffraction limit ($\lambda/2$).

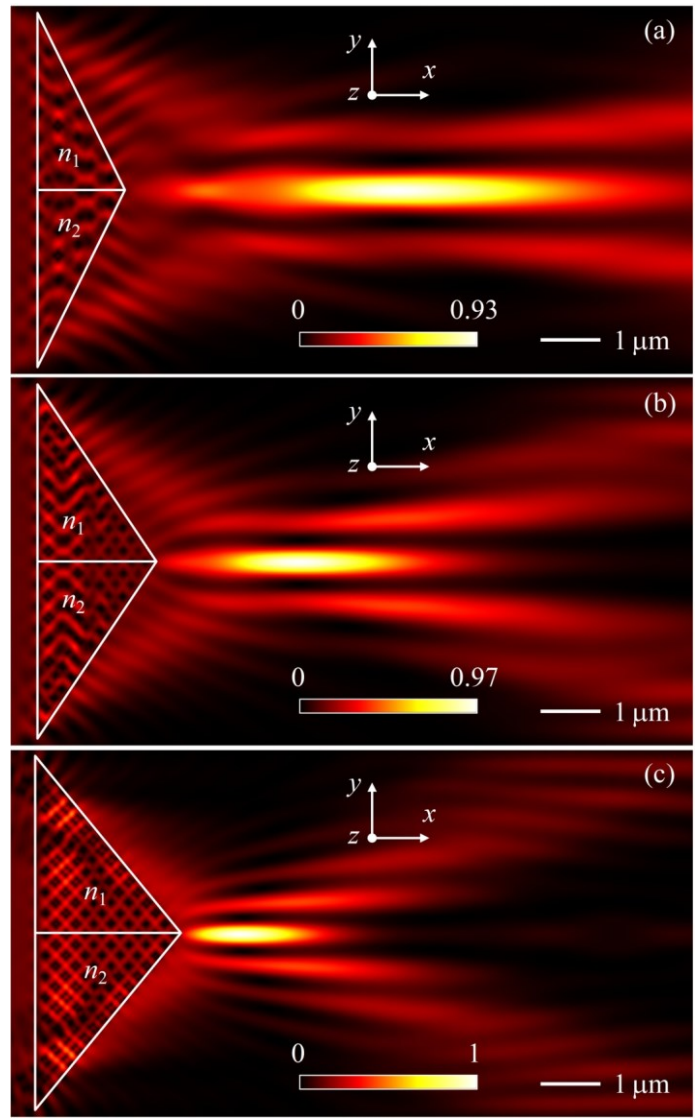

**Figure 2.** Normalized intensity distributions of the photonic nanojets formed by conventional triangular micro-prisms at (a) $h = 1.5$ μm, (b) $h = 2$ μm, and (c) $h = 2.5$ μm. The refractive index of triangular micro-prisms is $n_1 = n_2 = 1.5$.

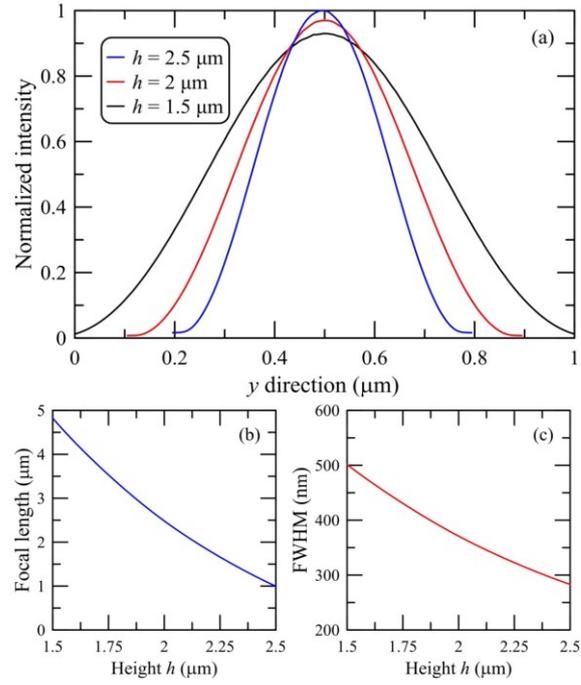

**Figure 3.** (a) Normalized intensity profiles of the photonic nanojets along $y$ direction for conventional triangular micro-prisms. (b) Focal length and (c) FWHM as a function of the height $h$ of the triangular micro-prisms.

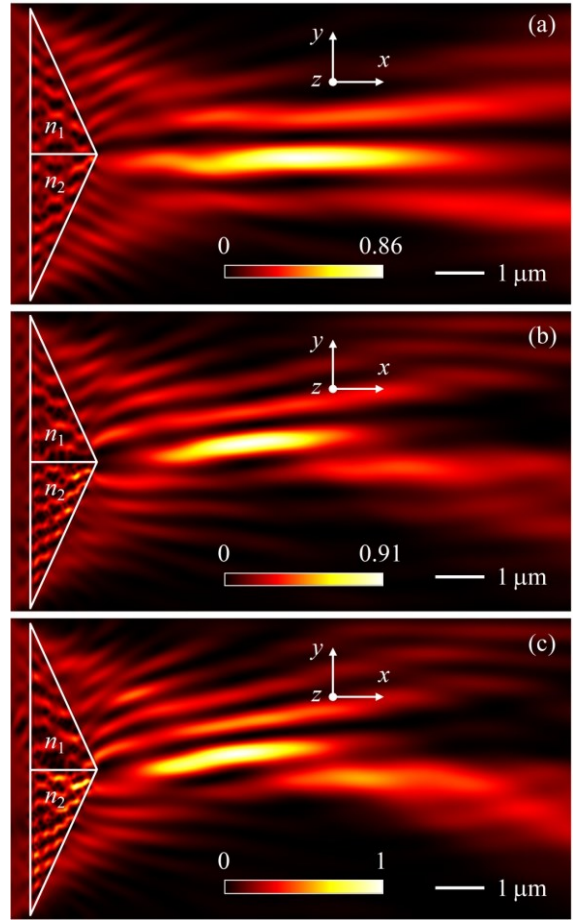

**Figure 4.** Normalized intensity distributions of the photonic hooks formed by triangular Janus micro-prisms at $n_1 = 1.5$, (a) $n_2 = 1.58$, (b) $n_2 = 1.88$, and (c) $n_2 = 1.95$. The height and width of triangular Janus micro-prisms are $h = 1.5$ μm and $w = 3$ μm.

In this article, we are interested in the possibility of forming curvilinear localized beams of the PH type. This can be achieved by introducing a refractive index gradient of the micro-prisms. Normalized intensity distributions of the PHs formed by triangular Janus micro-prisms with different refractive index contrasts are shown in Figure 4. The geometrical parameters of micro-prisms are $h$ = 1.5 μm and $w$ = 3 μm. The scattering of electromagnetic waves by the triangular micro-prisms leads to formation of the PH on the shadow-side of the micro-prisms. The length of the PH decreases as the refractive index contrast increases. Also, normalized intensity profiles of the PHs along $y$ direction for triangular mesoscale Janus prisms at different refractive index $n_2$, focal length, FWHM and bending angle [34] are shown in Figure 5. The dependencies of the key parameters of the PH are presented against the refractive index contrast of the two triangular micro-prisms. It can be clearly seen that an increase in the optical contrast of the materials of the two micro-prisms leads to a decrease in the FWHM and focal length, but to an increase in the curvature (bending angle $\theta$ – see Figure 1) of the PH.

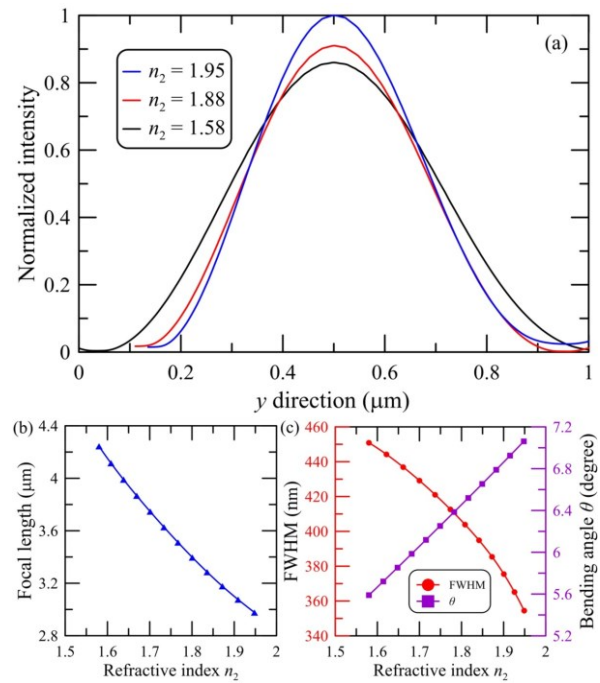

**Figure 5.** (a) Normalized intensity profiles of the photonic hooks along $y$ direction for triangular Janus micro-prisms at different refractive index $n_2$. (b) Focal length, (c) FWHM and bending angle as a function of the refractive index $n_2$ of the triangular micro-prisms

In the scheme of PH formation under consideration, there is one more additional degree of freedom, which makes it possible to control the characteristics of a localized structured electromagnetic flow. This is the space $d$ between two conjugate micro-prisms. We have used this additional parameter to modulate the PH bending angle. Normalized intensity distributions of the PHs formed by triangular Janus prisms at different space and corresponding key parameters of the PHs are shown below in Figures 6 and 7. The length of the PH increases as the space $d$ increases. Direct comparison of the results presented in Figures 5 and 7 demonstrates a fundamental difference in the key characteristics of the generated PHs. With an increase in the space between the conjugate micro-prisms with the remaining parameters of the problem, the focal length increases nonlinearly. The same trend is also observed for the minimum beam width in Figure 7(c). At the same time, the bending angle decreases as the space between the two micro-prisms increases. Note that the distance $d$ between the prisms in perspective can be used for the flow of environmental material and the analysis of nanoparticles in the region of the PH. But these studies are planned to be carried out in future works.

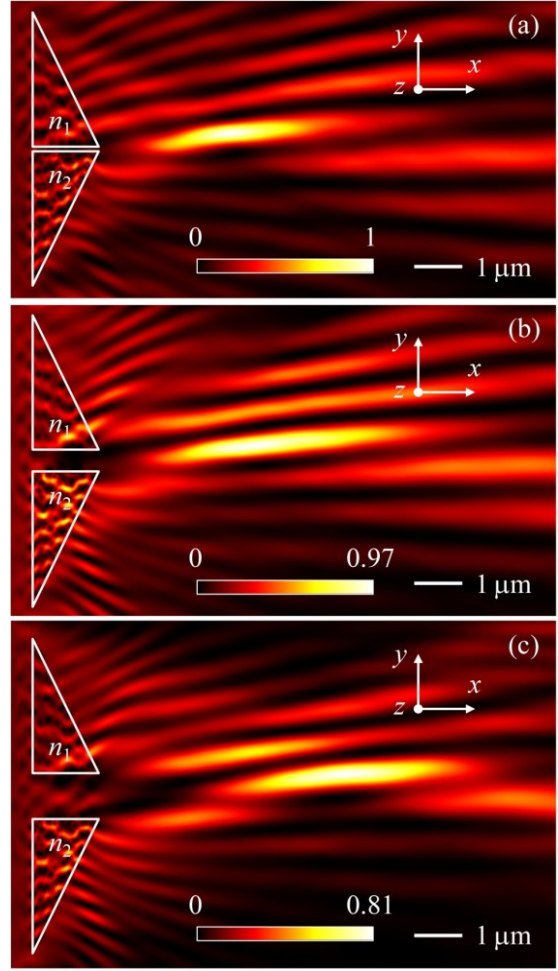

**Figure 6.** Normalized intensity distributions of the photonic hooks formed by triangular Janus micro-prisms at (a) $d$ = 100 nm, (b) $d$ = 500 nm, and (c) $d$ = 1000 nm. The refractive indices of triangular Janus micro-prisms are $n_1$ = 1.5 and $n_2$ = 1.88.

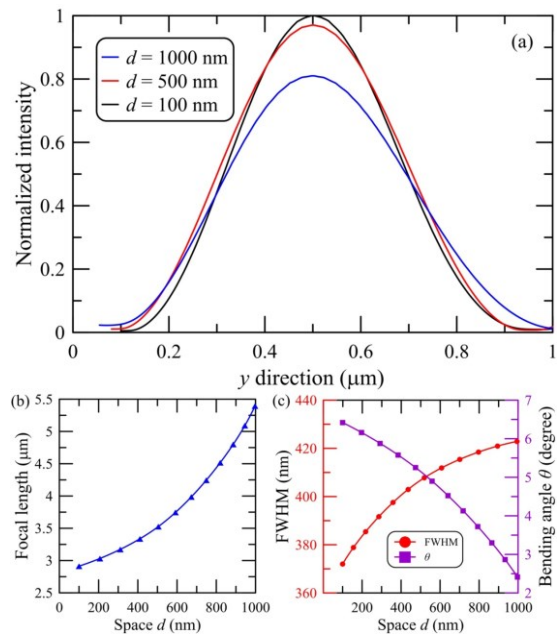

**Figure 7.** (a) Normalized intensity profiles of the photonic hooks along $y$ direction for triangular Janus micro-prisms at different spaces $d$. (b) Focal length, (c) FWHM and bending angle as a function of the space $d$ of the triangular micro-prisms.

To understand the physical process, Figure 8 shows Poynting vectors and energy flow streamlines for triangular micro-prisms at $n_1 = n_2 = 1.5$ (see Figure 2), $n_1 = 1.5$, $n_2 = 1.88$ (see Figure 4), and $n_1 = 1.5$, $n_2 = 1.88$, $d = 500$ nm (see Figure 6). In the case of the PJ formation, the subwavelength vortices inside the micro-prism are located symmetrically with respect to the axis of symmetry of the Janus particle. The introduction of optical contrast between conjugated prisms leads to a spatial redistribution of vortices. Accordingly, a curvature of the electromagnetic energy flux is generated behind the shadow part of the Janus particle, which is similar to square Janus particle consisting of two diagonally conjugate micro-prisms [28]. When the micro-prisms are separated by the space $d$, the energy flux is responsible for the curvature of the localized beam due to the influence of a part of the energy passing in the space between the two micro-prisms. This leads to a decrease in the banding angle of the beam as a whole. Moreover, an increase in the space $d$ also elongates the length, scatter and direction of the PH over a wider range. The use of the two triangular micro-prisms leads to a dependence of the PH length and PH curvature on the refractive index contrast and the space between the two micro-prisms.

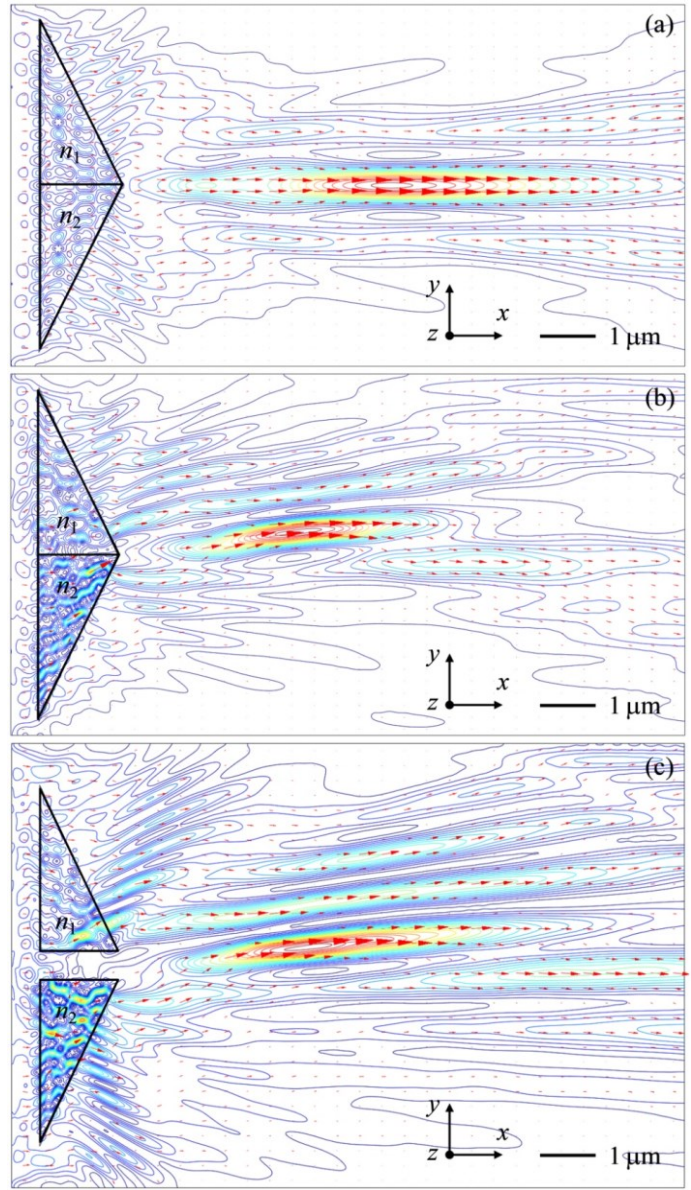

**Figure 8.** Poynting vectors and energy flow streamlines for triangular Janus micro-prisms at (a) $n_1 = n_2 = 1.5$, (b) $n_1 = 1.5$, $n_2 = 1.88$, and (c) $n_1 = 1.5$, $n_2 = 1.88$, $d = 500$ nm. The height and width of triangular Janus micro-prisms are $h = 1.5$ μm and $w = 3$ μm.

## 4. Conclusions

In summary, the scattering of electromagnetic waves by triangular Janus micro-prisms has been studied in order to demonstrate the possibility of long high-intensity PH formation. The relationship of the refractive index contrast and the space between the two micro-prisms is found to form the PH with subwavelength waist and bending angle on the shadow side of the Janus micro-prisms. The space between the two micro-prisms leads to an improvement in the characteristics of the PHs. It was shown that the triangular Janus micro-prisms make possible to focus optical beam in free space into the PH with waist smaller than the scalar diffraction limit. In particular, the triangular Janus micro-prism with $n_1 = 1.5$, $n_2 = 1.95$ forms a PH with a FWHM of 353 nm and a banding angle of 7°. Due to the asymmetric vortexes of intensity distributions, the long PH of 4.36 μm is obtained by the triangular Janus micro-prism at $n_1 = 1.5$, $n_2 = 1.88$, and $d = 500$ nm. By changing the space between the two micro-prisms, the PH length and the PH bending angle are efficiently modulated. From a practical point of view, a triangular Janus micro-prism can be fabricated by several modern technologies [39]. The effects of the asymmetric optical energy flow may apply many interesting applications such as nanoscopy in cell biology and nanoparticle trapping in the light-analyte interaction [40] procedures, in lab-on-a-chip microfluidics and microchannels systems [41,42], etc.

**Author Contributions:** Conceptualization, W.Y.C., C.Y.L. and I.V.M.; methodology, O.V.M. and I.V.M.; software, Y.K.H.; validation, O.V.M., W.Y.C. and C.Y.L.; formal analysis, W.Y.C.; investigation, O.V.M. and I.V.M.; resources, Y.K.H.; data curation, W.Y.C. and Y.K.H.; writing—original draft preparation, C.Y.L., O.V.M. and I.V.M.; writing—review and editing, W.Y.C., O.V.M. and C.Y.L.; visualization, Y.K.H. and C.Y.L.; supervision, I.V.M. and O.V.M.; project administration, O.V.M.; funding acquisition, C.Y.L. All authors have read and agreed to the published version of the manuscript.

**Funding:** This research was funded by National Science and Technology Council of Taiwan, grant number NSTC 111-2221-E-A49-102-MY2.

**Acknowledgments:** I.V.M. and O.V.M. acknowledge the Tomsk Polytechnic University Competitiveness Enhancement Program.

**Conflicts of Interest:** The authors declare no conflict of interest.